\documentclass[journal=jacsat,manuscript=letter, layout=twocolumn]{achemso}

\usepackage[version=3]{mhchem} 
\usepackage[utf8]{inputenc}
\usepackage[T1]{fontenc}  
\usepackage{pifont}       
\usepackage{graphicx}
\usepackage[dvipsnames]{xcolor}
\usepackage[colorlinks=true, linkcolor=RoyalBlue, citecolor=RoyalBlue]{hyperref}
\usepackage[per-mode=symbol, separate-uncertainty]{siunitx}
\usepackage{chemformula}
\usepackage[noabbrev]{cleveref}
\usepackage{nicefrac}
\usepackage{float}
\usepackage{wrapfig}
\usepackage{amssymb}
\usepackage{gensymb}
\usepackage{caption}
\usepackage{subcaption}
\usepackage{multicol}
\usepackage{enumitem}
\usepackage{natmove}
\usepackage{setspace}
\usepackage{comment}
\usepackage[normalem]{ulem}

\author{Viola Zeller}
\email{viola.zeller@ur.de}
\affiliation{Institut für Experimentelle und Angewandte Physik, Universität Regensburg, D-93040 Regensburg, Germany}
\author{Nadine Mundigl}
\affiliation{Institut für Experimentelle und Angewandte Physik, Universität Regensburg, D-93040 Regensburg, Germany}
\author{Paulo E. Faria~Junior}
\affiliation{Institut für Theoretische Physik, Universität Regensburg, D-93040 Regensburg, Germany}
\author{Jaroslav Fabian}
\affiliation{Institut für Theoretische Physik, Universität Regensburg, D-93040 Regensburg, Germany}
\author{Christian Schüller}
\affiliation{Institut für Experimentelle und Angewandte Physik, Universität Regensburg, D-93040 Regensburg, Germany}
\author{Dominique Bougeard}
\email{dominique.bougeard@ur.de}
\affiliation{Institut für Experimentelle und Angewandte Physik, Universität Regensburg, D-93040 Regensburg, Germany}

\title{Wurtzite quantum wires with strong spatial confinement: polarization anisotropies in single wire spectroscopy}

\abbreviations{IR,NMR,UV}
\keywords{GaAs quantum wires, wurtzite, polarization}

\begin{document}

\begin{abstract}
We report GaAs/AlGaAs nanowires in the one-dimensional (1D) quantum limit. The ultrathin wurtzite GaAs cores between 20-40\,nm induce large confinement energies of several tens of meV, allowing us to experimentally resolve up to four well separated subband excitations in microphotoluminescence spectroscopy. 
Our detailed experimental and theoretical polarization-resolved study reveals a strong diameter-dependent anisotropy of these transitions: We demonstrate that the polarization of the detected photoluminescence is governed by the symmetry of the wurtzite 1D quantum wire subbands on the one hand, but also by the dielectric mismatch of the wires with the surrounding material on the other hand. The latter effect leads to a strong attenuation of perpendicularly polarized light in thin dielectric wires, making the thickness of the AlGaAs shell an important factor in the observed polarization behavior. Including the dielectric mismatch to our k.p-based simulated polarization-resolved spectra of purely wurtzite GaAs quantum wires, we find an excellent agreement between experiment and theory.
\end{abstract}

\section{Introduction}
One dimensional (1D) quantum wires stand out amongst spatially quantum-confined solid-state structures because of unique quantized properties such as fully ballistic transport \cite{VanWees1988}, efficient light-matter interaction with particular relevance for the realization of lasers \cite{Kapon1989, Asada1985}, strong manifestations of correlation, \cite{Deshpande2010Nature} or new quasiparticle states \cite{Sau2010} and dispersion effects \cite{Rashba2012} in spin-orbit coupled systems. At the same time, the experimental exploration of these rich physics is challenged by the availability of suitable solid-state structures. One challenge is the realization of wires with confinement energies large enough to operate experiments in the 1D quantum limit, in particular for optical spectroscopy. In this regard, catalytically self-assembled ultrathin GaAs nanowires \cite{Dirnberger2019_APL, Vainorius2016, Loitsch2015, Corfdir2018} with very pure crystal phases \cite{Furthmeier2014} have recently opened new perspectives.
In this letter, we report 1D semiconductor quantum wires revealing up to four well separated subband transitions in single wire photoluminescence (PL) with several tens of meV confinement energies. Our core/shell nanowires feature ultrathin GaAs cores below 40 nm and an AlGaAs shell for which we have varied the thickness. The experimental polarization-resolved spectra and observed confinement energies match very well with our k.p simulations for purely wurtzite nanowires, including band-mixing in the valence bands. Our experimental and theoretical study also clearly demonstrates the importance of the AlGaAs shell thickness in single wire polarization-resolved studies: The \textit{intrinsic} electronic polarization selection rules are overlayed by \textit{extrinsic}, classical polarization anisotropies, resulting from the dielectric mismatch between the GaAs/AlGaAs nanowire and its environment. We find that AlGaAs shell thicknesses of several hundred nanometers are required to experimentally observe the pure wurtzite electronic 1D quantum wire polarization behavior for several quantum confined subbands.

\section{Experimental and theoretical details}
\subsection{Nanowire growth}
\begin{figure}
    \centering
    \includegraphics[]{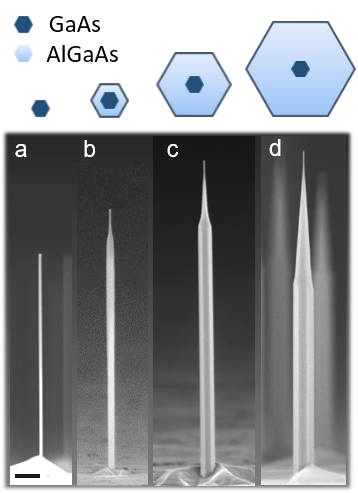} 
    \caption{Representative scanning electron micrographs of quantum wires from four different growth runs. Pure GaAs core without AlGaAs shell (a) and with different AlGaAs shell thicknesses (b-d). The total diameter of the wires amounts to 30\,nm (a), 70\,nm (b), 160\,nm (c) and 220\,nm (d) with lengths ranging from 2\,µm (a) to 3\,µm (d). Scale bar is 250\,nm. \label{fig1}}
\end{figure}

The investigated core/shell wurtzite GaAs quantum wires were grown by solid source molecular-beam epitaxy (MBE) on semi-insulating GaAs(111)\textsubscript{B} substrates via Au-catalyzed vapour-liquid-solid growth.
Prior to growth, the substrates were covered with a very thin Au-layer of nominally 0.05\,\si{\angstrom} which forms Au/Ga-droplets of $\sim$\,30\,nm diameter. 
The nanowire growth procedure was separated into two steps: 
In the first step, we synthesized the ultrathin GaAs cores at a substrate temperature of 540\,°C for 95\,min, a Ga rate of 0.7\,\si{\angstrom}/s and an As\textsubscript{4}/Ga ratio of 2.
This step yields vertically grown wires with high wurtzite crystal phase purity\cite{Dirnberger2019_APL, Dirnberger2019_Nanoletters, Furthmeier2014}, an uniform length of 2\,µm average and an average diameter of 30\,nm ($\pm$\,10\,nm), thin enough to observe 1D spatial quantum confinement effects \cite{Dirnberger2019_APL}. 
In the second step, the growth temperature was lowered to 460\,°C while the As\textsubscript{4}/Ga ratio was increased to $\sim$\,4, enabling a lateral overgrowth of the ultrathin GaAs core with an Al\textsubscript{0.38}Ga\textsubscript{0.62}As shell. 
In this way we fabricated several wafers with different AlGaAs shell thicknesses.
Figure \hyperref[fig1]{1} shows scanning electron micrographs (SEM) of as-grown nanowires without shell (a) and with various shell thicknesses corresponding to a lateral overgrowth time of 13\,min (b), 32\,min (c) and 63\,min (d). 
All nanowire shells were finally capped with nominally 5\,nm GaAs to prevent the oxidation of the AlGaAs.
We produced nanowires with total core/shell diameters ranging from 50\,nm to 230\,nm as determined by SEM, which we refer to as \textit{dielectric} diameter in the following, as the GaAs core and the Al\textsubscript{0.38}Ga\textsubscript{0.62}As shell have similar dielectric constants.
The larger bandgap of the Al\textsubscript{0.38}Ga\textsubscript{0.62}As shell confines all photocarriers to the GaAs core and suppresses non-radiative recombination caused by surface states, which is why we obtain a sufficient radiative efficiency in the photoluminescence (PL) even for the thinnest wires.

\subsection{Optical spectroscopy}
For PL experiments, the nanowires were randomly dispersed onto $300\,\text{nm}$ thermal SiO\textsubscript{2} formed on p-doped silicon chips by carefully brushing the growth substrate over the chip surface.
In our confocal microphotoluminescence (µ-PL) setup, the samples were excited by a pulsed semiconductor diode laser in a He-flow cryostat at 4.2\,K. 
We used two different PicoQuant lasers with peak energies of 1.8\,eV (690\,nm) or 3.1\,eV (405\,nm), $\sim$\,50\,ps pulse duration and a variable repetition frequency.
The laser beam was focused onto the sample through a 100x microscope objective with a numerical aperture of NA = 0.8, resulting in a spot diameter of $\sim$\,1\,µm, hence allowing us to address single nanowires lying on SiO\textsubscript{2}.
The PL signal was collected through the same objective and focused onto the entrance slit of a 75\,cm spectrometer, dispersed by a 150\,mm\textsuperscript{-1} grating and detected with a charge coupled device (CCD) camera. 
The exciting laser light was cut off by a suitable long-pass filter.

\subsection{Theoretical modeling}
The conduction and valence band states of the pure GaAs wurtzite nanowires are calculated via the k.p framework, within the formalism of the envelope function approximation and the plane wave expansion to treat the hexagonal quantum confinement. The valence band k.p Hamiltonian incorporates band-mixing effects of heavy, light, and crystal field split-off holes, which are crucial to properly model the intrinsic polarization anisotropies of optical spectra\cite{Wang2001, Persson2004PRB, Maslov2005PRB, FariaJunior2014JAP}. The conduction band is treated with a parabolic dispersion. Specific details of the k.p-based calculations can be found in Refs.~\cite{Campos2018PRB,Dirnberger2019_APL,Meier2021}. We use the valence band parameters and the electron effective masses given in Ref.~\cite{Cheiwchanchamnangij2011} (from the GW-LDA data set, without including the spin splitting terms in the valence band).

The optical signatures of the different subband excitations in the nanowires are investigated by calculating the absorption spectra\cite{Chuang1995book,Cardona2005book}
\begin{equation}
\alpha^a(\hbar\omega)=C_0\underset{c,v,k_z}{\sum} \left| p^a_{cvk_z} \right|^2 \, F_{cvk_z} \, \delta\left(\hbar\omega_{cvk_z}\!-\!\hbar\omega\right), 
\end{equation}
in which the superindex $a$ refers to the light polarization along $x$ (perpendicular) or $z$ (parallel to the nanowire axis) direction, the index $c(v)$ labels the conduction (valence) subbands, $k_z$ is the wave vector along the nanowire axis, $p^a_{cvk_z} = \left\langle c,k_z \left| \vec{p} \cdot {\hat{a}} \right| v,k_z \right\rangle$ is the dipole matrix element for the polarization of the light $\hat{a}$ which incorporates the valence band mixing and anisotropic Kane parameters\cite{FariaJunior2014JAP,FariaJunior2017PRB}, $F_{cvk_z} = \left(f_{c,k_z}-f_{v,k_z}\right)$, $f_{c(v),k_z}$ is the Fermi-Dirac distribution for the electron occupancy in the conduction (valence) subbands, $\hbar$ is the Planck's constant, $\omega_{cvk_z}$ is the interband transition frequency and $\delta$ is the Dirac delta function. The constant $C_0$ is $C_0 = 4\pi^2 e^2/(c_l n_r \varepsilon_0 m_0^2\omega \Omega)$, in which $e$ is the electron charge, $c_l$ is the speed of light, $n_r$ is the refractive index of the material, $\varepsilon_0$ is the vacuum permittivity, $m_0$ is the free electron mass, and $\Omega$ is the nanowire volume. In our calculations, we assume undoped systems at nearly zero temperature, so that $\left(f_{c\vec{k}}-f_{v\vec{k}}\right) = 1.$\cite{Kishore2010PRB} To include broadening effects due to finite carriers' lifetimes\cite{Chuang1995book,Chow1999book}, we replace the Dirac delta function by a Lorentzian function with a full width at half-maximum of 5 meV.

\section{Results and Discussion}
\subsection{Observation of wurtzite subbands}
\begin{figure*}
    \centering
	\includegraphics[]{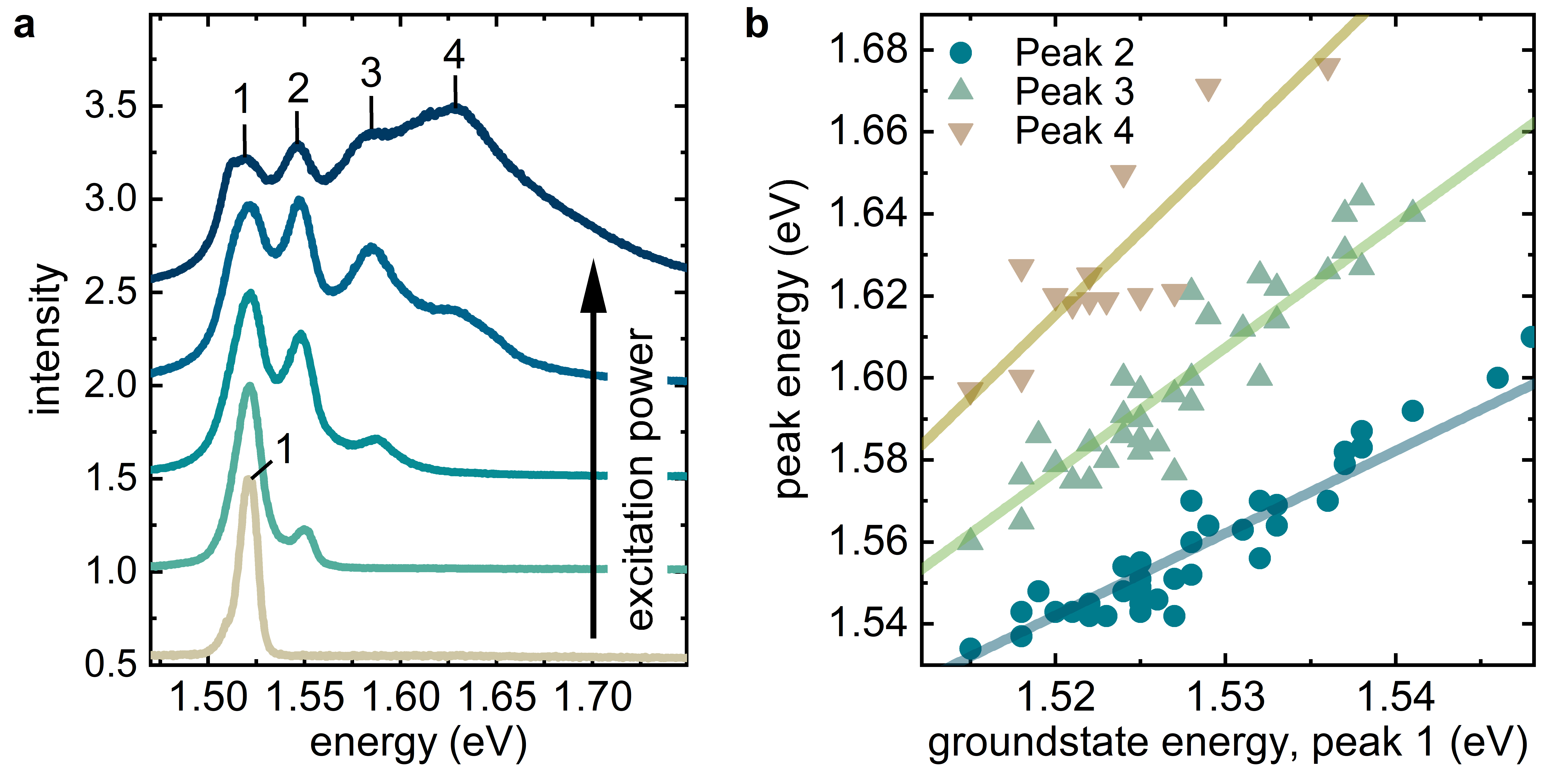}
	\caption{Observation of subband transitions in wurtzite quantum wires. (a) Spectra of a single quantum wire at increasing excitation intensities. With higher excitation, the subbands gradually fill up and appear as discernible peaks in the PL spectrum. The individual spectra are normalized to their maximum. (b) Subband energies of the investigated nanowires at high excitation power as a function of the ground state emission energy (at low excitation power). The linear dependence between the energies are indicative of subband transitions. The lines are guides for the eye.}
	\label{fig2}
\end{figure*}

We have investigated several tens of single quantum wires with different core and shell diameters. 
These wires were optically pre-selected to display a defect-free emission, which is correlated with a high crystal phase purity. \cite{Dirnberger2019_Nanoletters, Furthmeier2014, Ahtapodov2012} A characteristic excitation power series of the PL emission of a single wire is shown in Figure \hyperref[fig2]{2a}. While at low excitation power the emission (light brown curve) only displays the single narrow PL peak representative of the ground state emission (labeled "1" in \hyperref[fig2]{2a}), up to four peaks arise in the spectrum as we gradually increase the excitation power, labeled "1" to "4". Note that the emission peaks 1-4 do not significantly change their energetic position with increasing excitation level.

In Fig. \hyperref[fig2]{2b}, we summarize the peak energies observed in 37 single wires for the up to four peaks which arise with increasing excitation power. These wires span core diameters between approximately 20-40\,nm. The figure displays the peak energies of the peaks 2-4 as a function of the energy of their corresponding ground state emission peak 1. As a consequence of increasing spatial quantum confinement, the ground state PL energy of peak 1 continuously increases from 1.515\,eV to 1.548\,eV for decreasing core diameters. Furthermore, for all wires, the peak energies for peaks 2-4 clearly increase steadily, along with the corresponding ground state energy. All three peak energies 2-4 follow a roughly linear evolution as a function of the ground state energy 1, indicated as a guide to the eye for each peak in Fig. \hyperref[fig2]{2b}. The slightly larger scattering of the peak energies of peaks 3 and 4 is a result of some of these peaks being broader than peaks 1 and 2, increasing the error on our peak value determination. The observed relationship of peaks 2-4 to peak 1 suggests that, like peak 1, the emission energies of peaks 2-4 are governed by the same core diameter dependence for all 37 studied quantum wires. At the same time, we have observed that for a given core diameter, the energies for peaks 1-4 are independent of the shell thickness, i.e. of the dielectric diameter of the wire. We interpret these observations as a strong indication for peaks 2-4 to represent emission peaks from excited quantum wire subbands, successively filled with increasing excitation power.
We emphasize that the 37 quantum wires shown in Fig. \hyperref[fig2]{2b} feature different dielectric diameters. As a consequence, all the different shapes shown in Fig. \hyperref[fig1]{1b-d} are present in \hyperref[fig2]{2b}. Hence, while each wire in principle geometrically represents a cavity, we exclude peaks 1-4 to represent Fabry-Pérot resonances: It would be highly unlikely to observe the behavior documented in \hyperref[fig2]{2b} for Fabry-Pérot resonances in 37 wires with different dielectric diameters, tip shapes and wire lengths.

\subsection{Polarization resolved analysis}
Being able to resolve the emission of excited quantum wire subbands, opens up the possibility to experimentally uncover the polarization behavior of quantum wires in a wurtzite crystal: In bulk wurtzite GaAs, the degeneracy between the heavy hole (hh) and light hole (lh) valence band is lifted. As a result, the PL emission of bulk nanowires has been shown to be dominated by transitions from the hh-band, inducing a preferred polarization perpendicular to the nanowire axis. \cite{Ahtapodov2012, Furthmeier2014, Ketterer2011, DeLuca2015, Chen2008, BaHoang2010} Adding spatial quantum confinement, we have recently discussed a realistic model of the energetic subband structure of GaAs wurtzite quantum wires in the context of intersubband transitions resolved by resonant inelastic light scattering \cite{Meier2021}.

\begin{figure}
    \centering
    \includegraphics[]{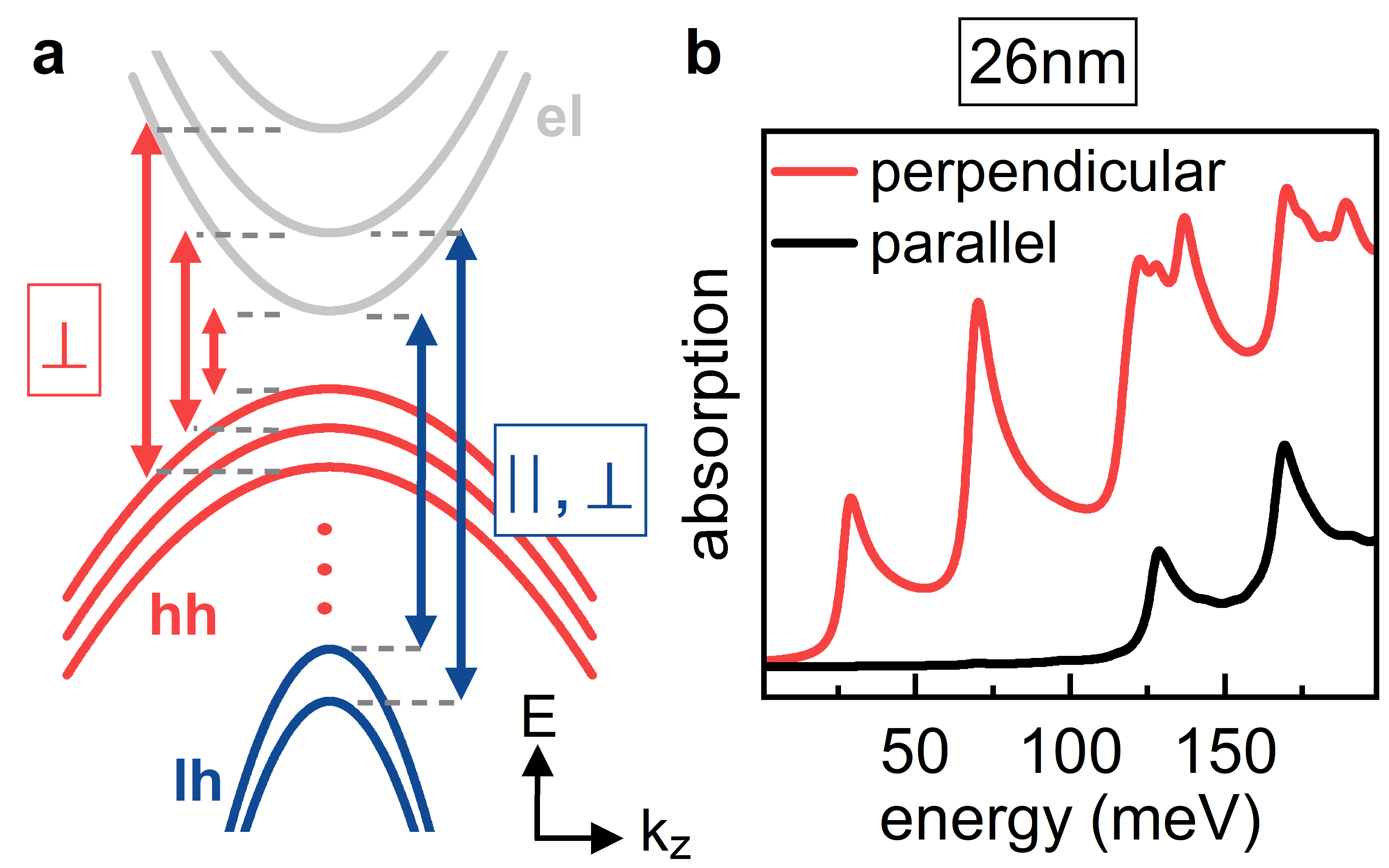}
	\caption{Polarization selection rules in 1D-confined wurtzite GaAs. (a) Sketch of the subband dispersion resulting from size quantization. The electronic conduction band (gray) and the non-degenerate heavy hole (red) and light hole (blue) valence bands are split into subbands. The arrows indicate the allowed dipole transitions with their respective allowed directions of polarization. (b) Calculated, polarization resolved absorption spectra of a wurtzite GaAs nanowire with 26\,nm core diameter, showing typical Van Hove singularities of a 1D quantum object. }
	\label{fig3}
\end{figure}

Figure \hyperref[fig3]{3a} sketches the electronic subband structure of a 1D quantum wire for the conduction (grey), the dominantly heavy hole (red) and the dominantly light hole (blue) bands.  
Note that dominantly hh transitions (red) only contribute to emission and absorption with perpendicular dipoles, while lh transitions (blue) contribute to both, perpendicular and parallel polarization. Hence, the character of the involved electronic transitions can be expected to strongly affect the polarization behavior of experimental emission and absorption spectra.

In Fig. \hyperref[fig3]{3b}, we show the realistic simulation of a polarization-resolved absorption spectrum for a GaAs quantum wire with 26\,nm core diameter. The spectrum is structured by the Van Hove singularities of the joint density of states (jDOS) for a 1D quantum object weighted by the allowed optical transitions. These singularities are resonances of the transitions highlighted by the arrows in \hyperref[fig3]{3a} and are offset with respect to the bulk wurtzite GaAs bandgap set here to $E_{\mathrm{gap}}=0$\,meV.  As can be seen in Fig. \hyperref[fig3]{3b}, our calculation for a wurtzite quantum wire with a 26 nm core diameter predicts the emission and absorption for energies up to 200\,meV above the bulk wurtzite bandgap to be dominated by perpendicular transition dipoles.

\begin{figure}
    \centering
	\includegraphics[]{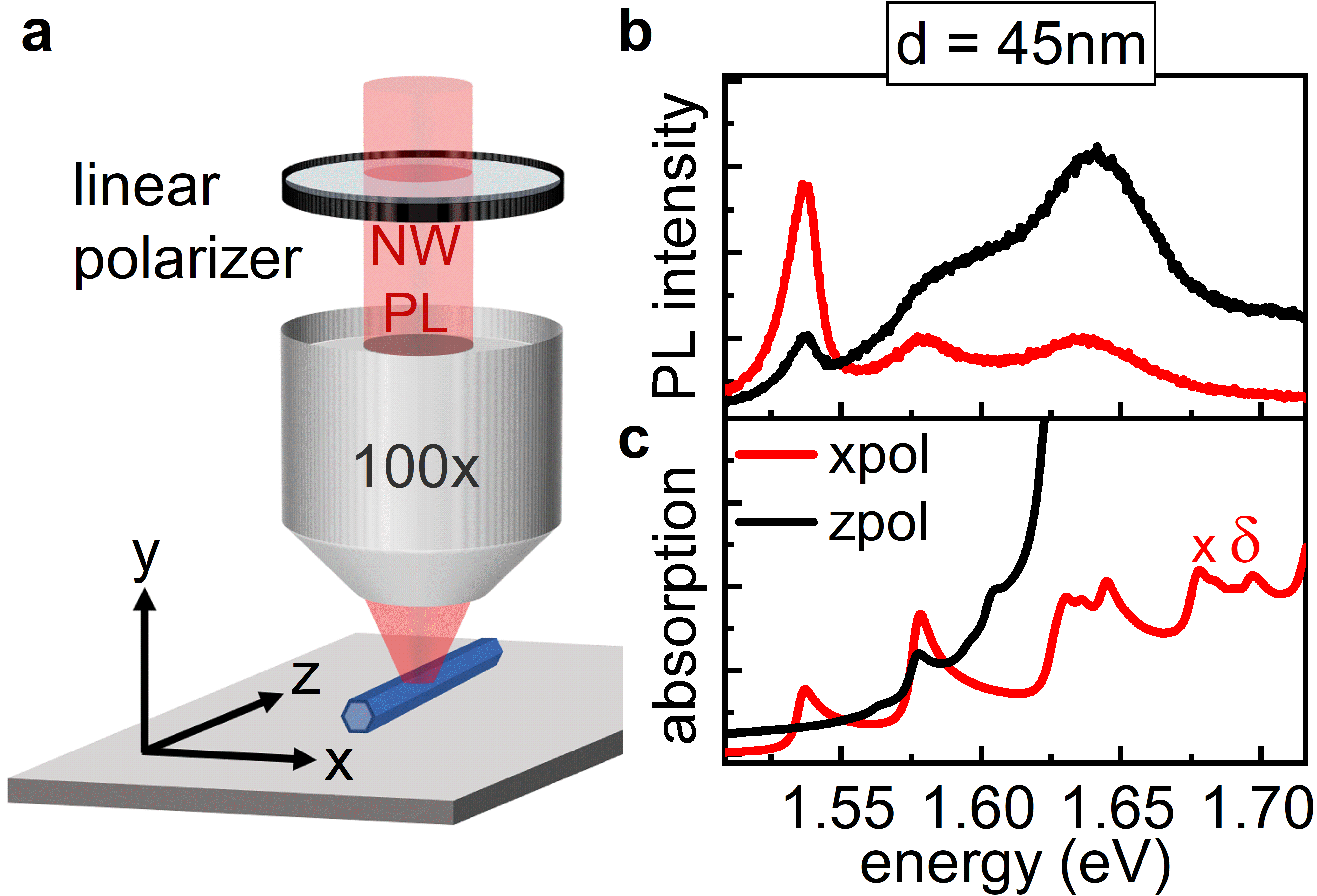}
	\caption{Polarization resolved analysis of thin quantum wires. (a) Simplified schematic of the measurement setup. (b) Experimental PL emission spectra analyzed perpendicular (xpol, red) and parallel (zpol, black) to the wire axis. (c) Calculated perpendicularly (red) and parallely (black) polarized absorption spectra. The x-polarized spectrum (red) was multiplied with the attenuation constant $\delta = 0.016$ according to our simulations. }
	\label{fig4}
\end{figure}

To test this prediction experimentally, we inserted a rotatable linear polarizer in the detection path and analyzed the polarization of the emitted light in x- and z-direction. A schematic of the setup is shown in Figure \hyperref[fig4]{4a}. We now discuss a series of quantum wires for which the energy of peaks 1-3 display excellent match with the first Van Hove singularities in Fig. \hyperref[fig3]{3b}. From this excellent match between the simulation and the energy positions of the PL peaks, we conclude that all the wires discussed in the following have a core diameter of 26\,nm.

In Fig. \hyperref[fig4]{4b} we start with the polarization-resolved spectra (x- and z-direction) of a quantum wire which features a comparatively thin dielectric diameter of 45\,nm, determined by SEM.
In accordance with the predicted dominant hh character from our simulation, the polarization of the ground state transition at 1.537\,eV clearly shows a preferential direction perpendicular to the wire axis in the PL spectrum.
Intriguingly, however, the dominant polarization of the emitted light changes from perpendicular to parallel at an energy of $\sim$\,1.55\,eV. The parallel polarized PL then clearly dominates the spectrum above $\sim$\,1.55\,eV. Hence, at first sight, the polarization in the experiment is in discrepancy with our simulation in Figure \hyperref[fig3]{3b}, which is solely based on interband transitions from the electronic 1D subband structure. Note, however and as mentioned before, that the energetic positions of peaks 1-3 in \hyperref[fig4]{4b} are in excellent match with \hyperref[fig3]{3b}.

This discrepancy in the polarization behavior between experiment and simulation, motivates us to consider possible \textit{extrinsic} impacts on the polarization behavior of the emission and absorption in nanowires and to explore whether they might add to or even outweigh the prediction of our \textit{intrinsic} electronic simulation in certain energy regions of the spectrum.
In this context, it has been discussed that the high aspect ratio, combined with the large refractive index contrast between the wire and its environment (vacuum), causes a strong polarization anisotropy. This dielectric mismatch effect favors the absorption and emission of parallel polarized light.
For instance, when a dielectric cylinder with the dielectric constant $\epsilon$ is placed into an external electric field $\mathbf{E_0}$, the component normal to the wire axis is essentially suppressed inside the wire: \cite{LANDAU198434, Cheiwchanchamnangij2011, Ruda2005, Wang2001, Chen2008}
\begin{equation}\label{E_perp}
    E_{\perp} = \frac{2 \epsilon_0}{\epsilon + \epsilon_0} E_{0\perp},
\end{equation}

\noindent where $\epsilon_0$ is the vacuum permittivity, $E_{\perp}$ and $E_{0\perp}$ the normal electric field components inside and outside the wire, respectively.
At the same time, the field component parallel to the wire is not reduced: $E_{\parallel} = E_{0\parallel}$.
Though eq. (\ref{E_perp}) was initially derived for static electric fields, it was shown to also be valid for electromagnetic waves, as long as their wavelength can be considered to be much larger than the diameter of the wire, $\lambda \gg d$.\cite{LANDAU198434, Cheiwchanchamnangij2011, Ruda2005, Wang2001}
Given that $E_{0\parallel} = E_{0\perp}$, the damping constant for the perpendicular field inside the wire amounts to

\begin{equation}\label{delta}
  \delta = \frac{I_x}{I_z} = \frac{|E_{\perp}|^2}{|E_{\parallel}|^2} = \Big| \frac{2 \epsilon_0}{\epsilon + \epsilon_0}\Big|^2 .
\end{equation}

\noindent Given that nanowires rarely fulfill the condition $\lambda \gg d$, we go beyond Eq. \ref{delta} by estimating the attenuation of the perpendicular electric field $\delta = I_x/I_y$ as a function of nanowire diameter by finite-element-method (FEM) based modeling (see \textcolor{blue}{Supplementary Information} for a detailed description).
For the wire discussed in Fig. \hyperref[fig4]{4b}, we find $\delta = 0.016$.
Fig. \hyperref[fig4]{4c} then displays again our calculated absorption spectra discussed in Fig. \hyperref[fig3]{3b}, but, in addition, the x-polarized spectrum has now been multiplied with $\delta = 0.016$, to take into account the dielectric mismatch effect for a wire with a diameter of 45\,nm. We have shifted the  absorption spectrum rigidly so that the first singularity coincides with the ground state PL peak of Fig. \hyperref[fig4]{4b}. Clearly, the simulated absorption spectrum Fig. \hyperref[fig4]{4c} predicts a preferred z-polarization above 1.55\,eV, while x-polarization dominates below this energy. As a consequence, the simulation is now in excellent agreement with the characteristic features of the experimental emission spectra: not only do the energetic positions of the first four PL peaks match with calculated electronic subband excitations, but also the switching from a dominant x-polarization to a dominant z-polarization around 1.55\,eV is correctly predicted. Note that we analyzed several nanowires with dielectric diameters around 45\,nm and observed this switching of the dominant polarization for all of them, even for core diameters smaller or larger than the discussed 26\,nm. Hence, we interpret the strong attenuation of the perpendicular transitions in these wires as a consequence of strong dielectric mismatch.

\begin{figure*}
    \centering
	\includegraphics[]{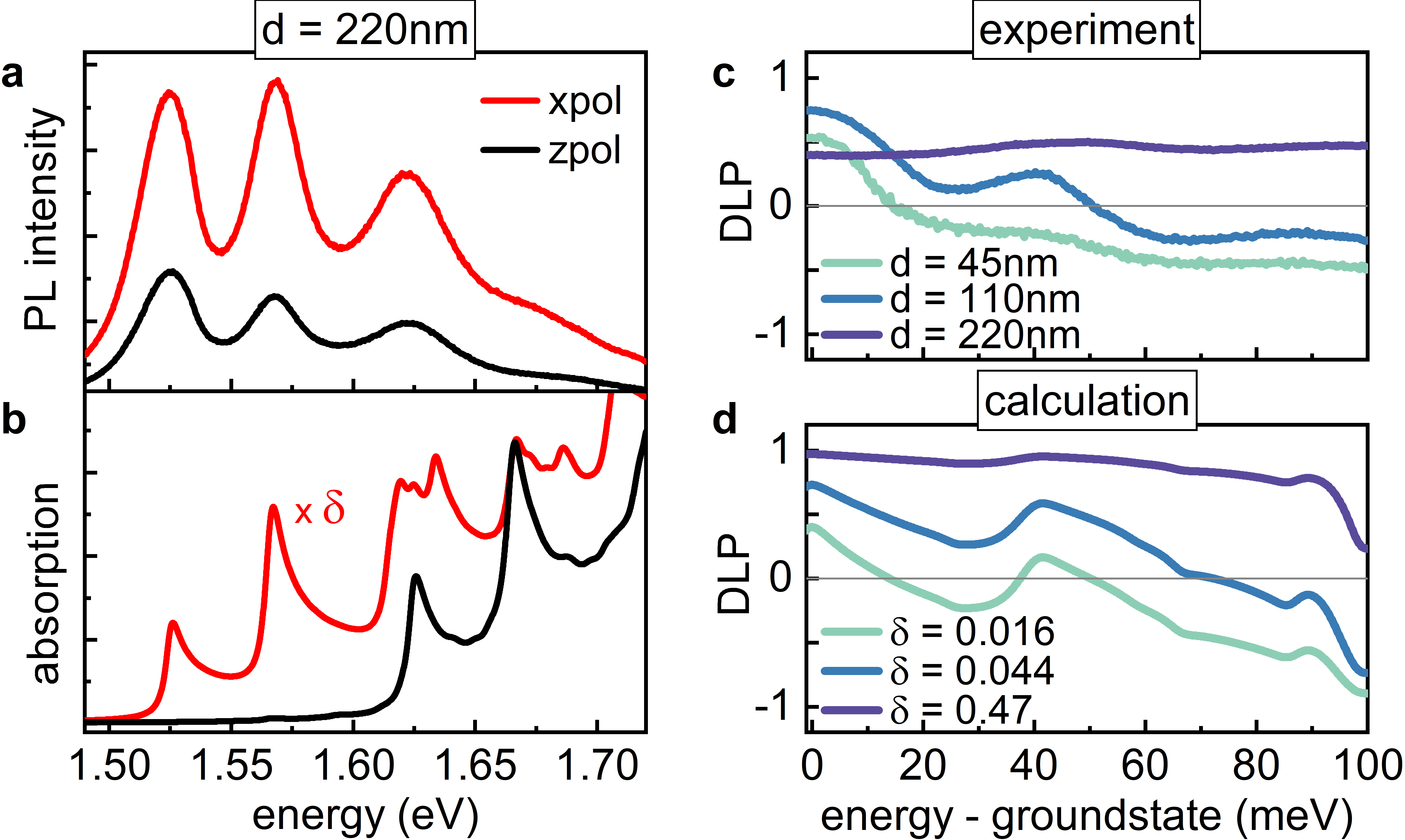}
	\caption{(a) Polarization resolved PL emission spectra of a quantum wire with a dielectric diameter of 220\,nm and a core diameter of 26\,nm. 
	(b) Corresponding calculated absorption spectra of a 26\,nm wire where the x-polarized spectrum was multiplied with $\delta=0.47$.
	(c) Energy dependent DLP curves of three wires with a 26\,nm core and different dielectric diameters of d = 45\,nm, d = 110\,nm and d = 220\,nm respectively.
	(d) Energy dependent DLP curves of the calculated absorption spectra with different $\delta$-values found from our FEM simulations for the corresponding diameters in (c).
    }
	\label{fig5}
\end{figure*}

Our model system of wurtzite GaAs quantum wires with variable shell thickness allows us to further specify the effect of the dielectric mismatch on the polarization behavior as a function of the dielectric diameter. In particular, the attenuation of the perpendicular electric field component should gradually be weakened when increasing the dielectric diameter. Thus, we now discuss the polarization-resolved spectra for a much thicker wire featuring a dielectric diameter of 220\,nm. At the same time, compared to Figure \hyperref[fig4]{4b}, we kept the GaAs quantum wire core diameter constant at 26\,nm. From our FEM simulations we calculate $\delta = 0.47$ for a dielectric diameter of 220\,nm. In Figure \hyperref[fig5]{5a}, we show the experimental polarization-resolved PL spectra and compare them to the corresponding, calculated absorption spectra, plotted in \hyperref[fig5]{5b}. Like in Fig. \hyperref[fig4]{4c}, the x-polarized simulated spectrum was multiplied by the attenuation factor, here $\delta = 0.47$. 
The three major PL peaks of the experiment agree very well with the first Van Hove singularities predicted for this 26\,nm quantum wire core. At the same time, contrary to the thinner 45\,nm dielectric diameter, this 220\,nm dielectric diameter reveals a preferred polarization perpendicular to the wire axis over the entire experimentally covered energy range, in excellent consistency with the simulated spectra. 

Note that the experimental parallel spectrum reveals two peaks in the energy range of 1.50\,eV - 1.60\,eV which are not predicted by the simulation in Fig. \hyperref[fig5]{5b}.
As we discuss in more detail in the \textcolor{blue}{Supplementary Information}, although our experimental configuration excludes the collection of perpendicularly y-polarized emitted light, we attribute these peaks to stem from y-polarized emitted light which is scattered into z-polarization primarily at the nanowire end facets. Given that the wave guiding of perpendicularly polarized light is more efficient for thicker dielectric wire diameters,  we expect this scattering effect to be more pronounced in our wire with 220\,nm diameter than in the 45\,nm wire. Hence, the two peaks are clearly visible in the Fig. \hyperref[fig5]{5a} in the 220\,nm wire, while their occurrence is less obvious in the 45\,nm wire in Fig. \hyperref[fig4]{4b}.

To demonstrate that the dominant modulation mechanism of the \textit{intrinsic} polarization-resolved emission of our quantum wires is the dielectric diameter-dependent attenuation, we finally compare the 45\,nm, the 220\,nm and a wire with an intermediate dielectric diameter thickness of 110\,nm in Fig. \hyperref[fig5]{5c} and \hyperref[fig5]{5d} (constant core diameter, 26\,nm). We plot the energy-dependent degree of linear polarization (DLP) defined as

\begin{equation}
 \mathrm{DLP} = \frac{(I_x - I_z)}{(I_x + I_z)} , 
\end{equation}
 so that the pure x-polarization of the light yields DLP = +1 and pure z-polarization DLP = -1.

Fig. \hyperref[fig5]{5c} shows the experimentally determined DLP for all three wires, as extracted from the polarization-resolved PL spectra. We compare these traces to the DLP extracted from our calculated absorption spectra shown in Fig. \hyperref[fig5]{5d}, for which the x-polarization was modulated with the attenuation $\delta$-value corresponding to each dielectric diameter. In order to align the energy scale, the energy of the ground state transition was set to zero for all curves.

The simulation quite precisely matches the energetic positions and magnitudes of the experimental DLP variations for all three wires, as prominently seen for the two distinct peak-like features for the 45\,nm and 110\,nm wires.
As can be seen in Fig. \hyperref[fig3]{3b}, for our representative example of 26\,nm quantum wire core diameter, the significant hh-lh splitting induced by wurtzite GaAs and by the quantum confinement leads to negligible parallel polarized contributions in absorption and emission in the energy window 0-100 meV considered in Fig. \hyperref[fig5]{5c,d}. Hence, wires with a dielectric diameter thick enough to display negligible attenuation, i.e. a $\delta$-factor close to 1, delivers a positive DLP, close to DLP=+1. From the comparison of Fig. \hyperref[fig5]{5c,d}, we see that this is still true even for $\delta=0.47$ in wires with a dielectric diameter of 220\,nm. As mentioned before, we attribute the discrepancy in absolute DLP values between experiment and simulation to scattering effects in the experiments, which are not included in our simulations. 

Although the parallel polarized contribution is negligible between 0-100\,meV in the electronic absorption and emission spectra, our simulation predicts that the lh-character - which produces parallel polarization - is non-zero. It actually increases from 0 towards 100\,meV depending on the strength of the dielectric mismatch effect. Remarkably, the electromagnetic phenomenon of the dielectric mismatch effect and its attenuation of $E_{\perp}$ allows to reveal this non-zero contribution of the parallel polarization experimentally. The DLP traces for the wires with 110\,nm and 4 \,nm show that the experimental visibility of the spectral structure of the parallel polarized contributions increases with decreasing dielectric wire diameter, i.e. with decreasing $\delta$-value.
This excellent match between simulations and experiments highlights that the three DLP spectra are the result of a subtle superposition of the polarization behavior imposed by the electronic transitions and the attenuation of the perpendicular field components due to the diameter-dependent dielectric mismatch effect.

\section{Conclusion}
In conclusion, our MBE grown core/shell nanowires reproducibly provide ultrathin wurtzite GaAs cores. Such cores ease experiments in the 1D quantum limit \cite{Kapon1989, Asada1985, Furthmeier2016, Deshpande2010Nature,Dirnberger2019_APL}, given their large confinement energies of several tens of meV. We found an excellent agreement between polarization-resolved photoluminescence spectra and k.p-based simulations for the excitation of up to four well-separated subband transitions. At the same time, while it does not modify the 1D quantum confinement of the GaAs core, we have revealed a significant role of the thickness of the AlGaAs shell in polarization-resolved spectroscopy of such 1D quantum wires. Our experiments and simulations show that for shells below thicknesses around 200\,nm, the dielectric mismatch effect will significantly overlay the intrinsic electronic 1D wurtzite selection rules. Comparatively large shell thicknesses above 200\,nm were required to observe the polarization behavior predicted by the electronic selection rules for up to four 1D subband excitations.

\begin{acknowledgement}
We acknowledge the financial support of the Deutsche Forschungsgemeinschaft through Project ID 422 31469 5032-SFB1277 (Subprojects A01, B05, B11).
\end{acknowledgement}

\bibliography{02_paper_polarization}

\end{document}